\documentclass[iop,apjl]{emulateapj}

\usepackage{apjfonts,ifthen,graphicx,times,amsmath,ifpdf}

\newcommand{\msun}{M$_\sun$}

\newcommand{\pdcflux}{\texttt{PDC\_FLUX}}
\newcommand{\sapflux}{\texttt{SAP\_FLUX}}
\newcommand{\sappdc}{\texttt{SAPPDC}}

\begin{document}

\title{Kepler Observations of Rapid Optical Variability in Active Galactic Nuclei}

\shorttitle{KEPLER OBSERVATIONS OF AGN}
\shortauthors{MUSHOTZKY ET AL}

\submitted{(Submitted 9 Sep 2011; Accepted 31 Oct 2011)}
 \journalinfo{(accepted by the Astrophysical Journal Letters)}

\author{
R.~F.~Mushotzky\altaffilmark{1,2,5},
R.~Edelson\altaffilmark{1},
W.~H.~Baumgartner\altaffilmark{2,3},
P.~Gandhi\altaffilmark{4}
}

\altaffiltext{1}{Department of Astronomy, University of
 Maryland College Park, College Park, MD 20742}
\altaffiltext{2}{NASA/Goddard Space Flight Center, Astrophysics
 Science Division, Greenbelt, MD 20771}
\altaffiltext{3}{Joint Center for Astrophysics, University of
 Maryland Baltimore County, Baltimore, MD 21250}
\altaffiltext{4}{Institute of Space and Astronomical Science, Japan
  Aerospace Exploration Agency, 3-1-1 Yoshinodai, chuo-ku, Sagamihara,
Kanagawa 252-5210, Japan}
\altaffiltext{5}{Corresponding author: richard@astro.umd.edu}

\begin{abstract}

Over three quarters in 2010--2011, Kepler monitored optical emission
from four active galactic nuclei (AGN) with $\sim$30 min sampling,
$>90$\% duty cycle and $\lesssim 0.1$\% repeatability.  These data
determined the AGN optical fluctuation power spectral density
functions (PSDs) over a wide range in temporal frequency.  Fits to
these PSDs yielded power law slopes of $-2.6$ to $-3.3$, much steeper
than typically seen in the X-rays.  We find evidence that individual
AGN exhibit intrinsically different PSD slopes.  The steep PSD fits
are a challenge to recent AGN variability models but seem consistent
with first order MRI theoretical calculations of accretion disk
fluctuations.

\end{abstract}

\keywords{Accretion, accretion disks --- Black hole physics ---
Galaxies: active --- Galaxies: Seyfert}

\section{Introduction}

The optical continuum from active galactic nuclei (AGN) is believed to
be dominated by emission from an accretion disk surrounding a
supermassive black hole and can be adequately modeled as radiation
from a simple Shakura-Sunyaev disk (Edelson \& Malkan 1986).  Because
this region is too small to image (except via gravitational lensing;
Kochanek 2004), indirect methods must be used to probe its structure
and physical conditions.  One of the best probes is provided by the
strong variability seen throughout the optical/ultraviolet/X-ray bands
in most AGN.  However, limitations with many ground-based optical
observations have made it difficult to obtain accurate, densely and
regularly sampled data sets covering the large range of timescales
necessary to constrain disk physics and search for characteristic
times which may be related to orbital, dynamic or other expected
timescales.  In particular, diurnal and weather related interruptions
can severely degrade the ground based sampling pattern and atmospheric
seeing introduces photometric errors that are much larger than the
Kepler uncertainties and often are as large as or larger than the
intrinsic short timescale optical source variability.  However ground
based data have sampled much longer timescales than are available in
the present Kepler data sets.

The natural timescales for a disk---light-crossing (t$_l$), dynamical
(t$_{dyn}$), and thermal (t$_{th}$) timescales---are set by the black
hole mass and the accretion processes (Frank, King \& Raine 2002). The
order of magnitude estimates for these timescales are: t$_l =
2.6$~M$_7$~$R_{100}$~hours, t$_{dyn} =
10$~M$_7$~$R_{100}$$^{3/2}$~days, and t$_{th} =
0.46$~M7~$R_{100}$$^{3/2}~\alpha_{0.01}$$^{-1}$~years, where M$_7$ is
the black hole mass in units of $10^7$~\msun, $R_{100}$ is the
emission distance in units of 100 times the Schwarzschild radius
$2GM/c^2$, and $\alpha_{0.01}$ is the Shakura-Sunyaev viscosity
parameter (Shakura \& Sunyaev 1973) divided by 100. For assumed
Eddington ratios of 0.01--0.1 and mass ranges of
$10^6$--$10^9$~\msun\ typical for AGN, these natural timescales range
from hours to years.  Previous data have been unable to constrain the
optical time variability over this wide range for any individual AGN.

The Kepler mission (Borucki et al. 2010) provides a solution to these
observational difficulties.  Kepler has been observing a $\sim$115
square degree region of sky, monitoring $\sim$165,000 sources every
29.4 minutes with unprecedented stability ($\lesssim 0.1$\% for a
15th magnitude source) and high duty cycle ($>90$\%) over a period of
years.  During Q6 (Quarter 6: UT 24 June--22 September 2010), Q7 (23
September--22 December 2010) and Q8 (22 December 2010--24 March 2011),
the Kepler target list included at least four variable AGN from our
guest observer program.  This paper reports initial results of Q6--Q8
(and in one instance Q4) observations of these Kepler AGN, focusing on
fluctuation power spectral density analysis.  The source selection,
data collection and reduction are given in Section 2, the time series
analysis and results are reported in Section 3, implications are
discussed in Section 4, and brief conclusions presented in Section 5.

\section{Data}

\subsection{Source Selection}

Because it lies at low galactic latitudes not systematically covered
by major extragalactic or AGN surveys the Kepler field ($\sim$0.3\% of
the sky) currently contains only a few catalogued
AGN\footnote{However, a portion of the Kepler field is covered by
  SDSS/SEGUE, \texttt{http://www.sdss.org/segue/}}. Targets must be
identified and windows chosen before Kepler data can be
downloaded. Thus we have undertaken extensive efforts to identify AGN
in the Kepler field. This started with a database search to find
previously identified AGN.  We then applied the method of Stocke et
al. (1983) to the Rosat all sky survey (RASS; Voges et al. 1999) to
select AGN candidates based on their X-ray to optical flux ratio. We
also used the 2MASS all sky survey catalog (Strutskie et al. 2006) to
identify AGN candidates based on infrared colors (Malkan 2004) and
association with a RASS source.

Table~\ref{table1} gives details of the Kepler AGN whose light curves
are presented in this paper, a sample of four variable AGN that Kepler
has been observing since Q6.  Of these four, only Zw~229$-$15 ($z =
0.0275$, Falco et al. 1999, Proust 1990) had been identified as an AGN
prior to the launch of Kepler. A recent reverberation mapping campaign
found it had an H$\beta$ lag of $\sim$4 days and estimated its black
hole mass at $\sim10^7$~\msun\ (Barth et al. 2011).  The other three
AGN in Table~\ref{table1} were all discovered as a result of the
search described above.  (The prefix ``KA'' is used to designate newly
identified Kepler AGN.)  Spectra of these three, plus ten other newly
discovered Kepler AGN are given in Edelson \& Malkan (2012).

\begin{deluxetable}{lrccrr}
\tablecaption{Kepler AGN Reference Information \label{table1}}
\tablewidth{\columnwidth}
\tablehead{
\colhead{Source Name} &\colhead{Kepler ID} &\colhead{RA(J2000)} &\colhead{Dec(J2000)} 
&\colhead{z} &\colhead{RASS} }
\startdata
Zw~229$-$15 & 6932990 & 19 05 26.0 & +42 27 40 & 0.028 & 0.450\\
KA 1925$+$50 & 12158940 & 19 25 02.2 & +50 43 14 & 0.067 & 0.170\\
KA 1858$+$48 & 11178007 & 18 58 01.1 & +48 50 23 & 0.079 & 0.210\\
KA 1904$+$37 & 2694185 & 19 04 58.7 & +37 55 41 & 0.089 & 0.023
\enddata
\tablecomments{Columns 1 and 2 give the source name (KA refers to
  newly discovered Kepler AGN first reported in this paper) and
  Kepler ID number, columns 3 and 4 give the position, column 5 the
  redshift, and column 6 the Rosat all sky survey (RASS) count rate in
  counts~s$^{-1}$.  }
\end{deluxetable}

\subsection{Kepler SAP Light Curves}

\begin{figure*}
  \begin{center}
    \ifpdf
      \resizebox{\textwidth}{!}{\includegraphics{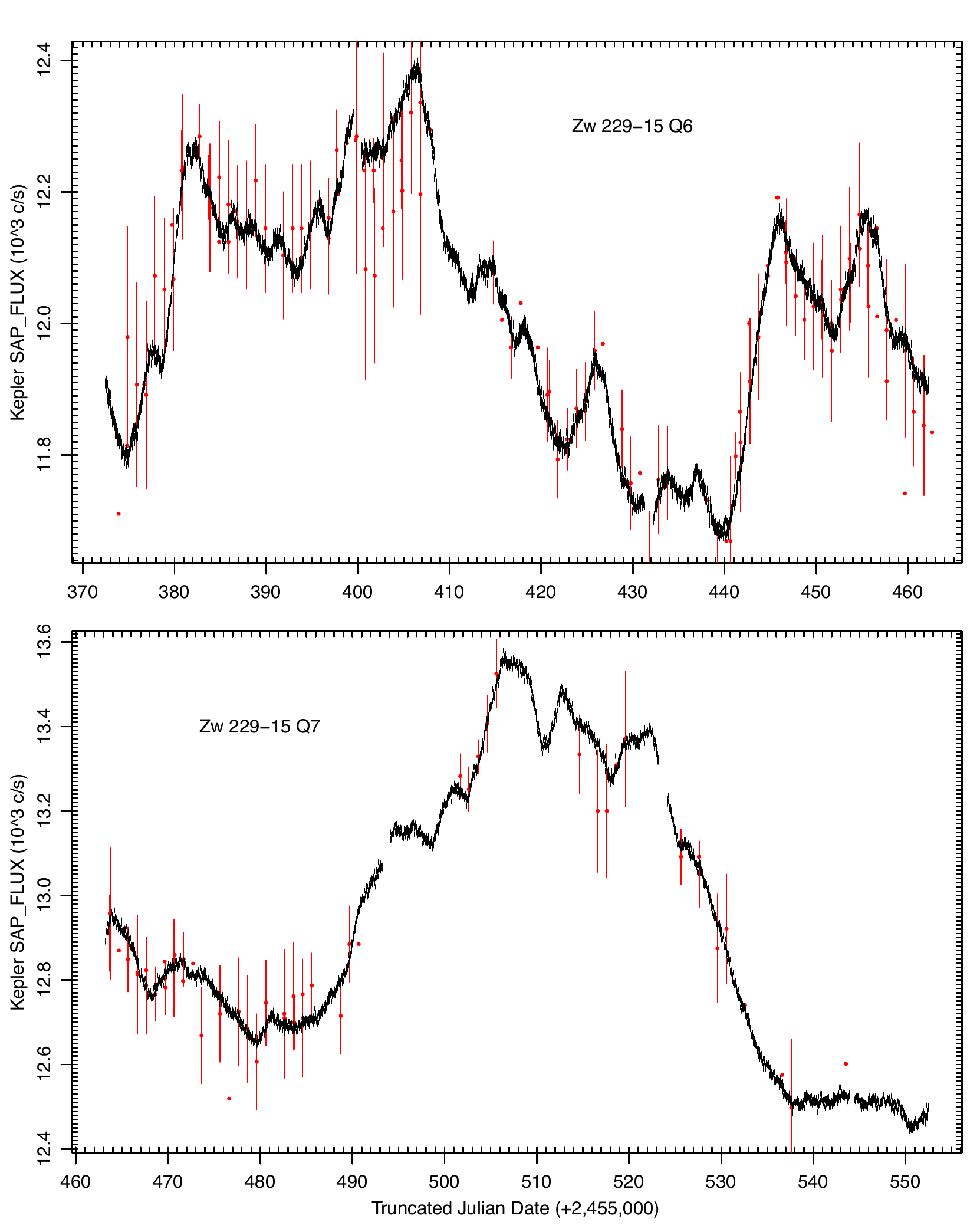}}
    \else
      \resizebox{\textwidth}{!}{\includegraphics{fig1.eps}}
    \fi
  \end{center}
\caption{Kepler Q6 (top) and Q7 (bottom) light curves of the narrow
  line Seyfert~1 galaxy Zw~229$-$15 (in black).  Each panel contains
  over 4,200 cadences, gathered one every $\sim$30 min, with a
  precision of $\lesssim$0.1\%.  A typical error bar is seen in the
  outlier at TJD $\sim$ 539.  There are monthly $\sim$1 day data
  download gaps (e.g., TJD $\sim$ 431 and 524), but the overall duty
  cycle is $>90$\%.  Note the $\sim$8\% flux discontinuity between Q6
  and Q7 as the quarterly spacecraft roll moves the source onto a
  different chip and a new SAP aperture is used.  Note also the
  excellent agreement with simultaneous ground based LAMP data (shown
  in red; Barth et al. 2011), scaled to account for different aperture
  sizes.
\label{figure1}}
\end{figure*}

\begin{figure*}
  \begin{center}
    \ifpdf
      \resizebox{\textwidth}{!}{\includegraphics{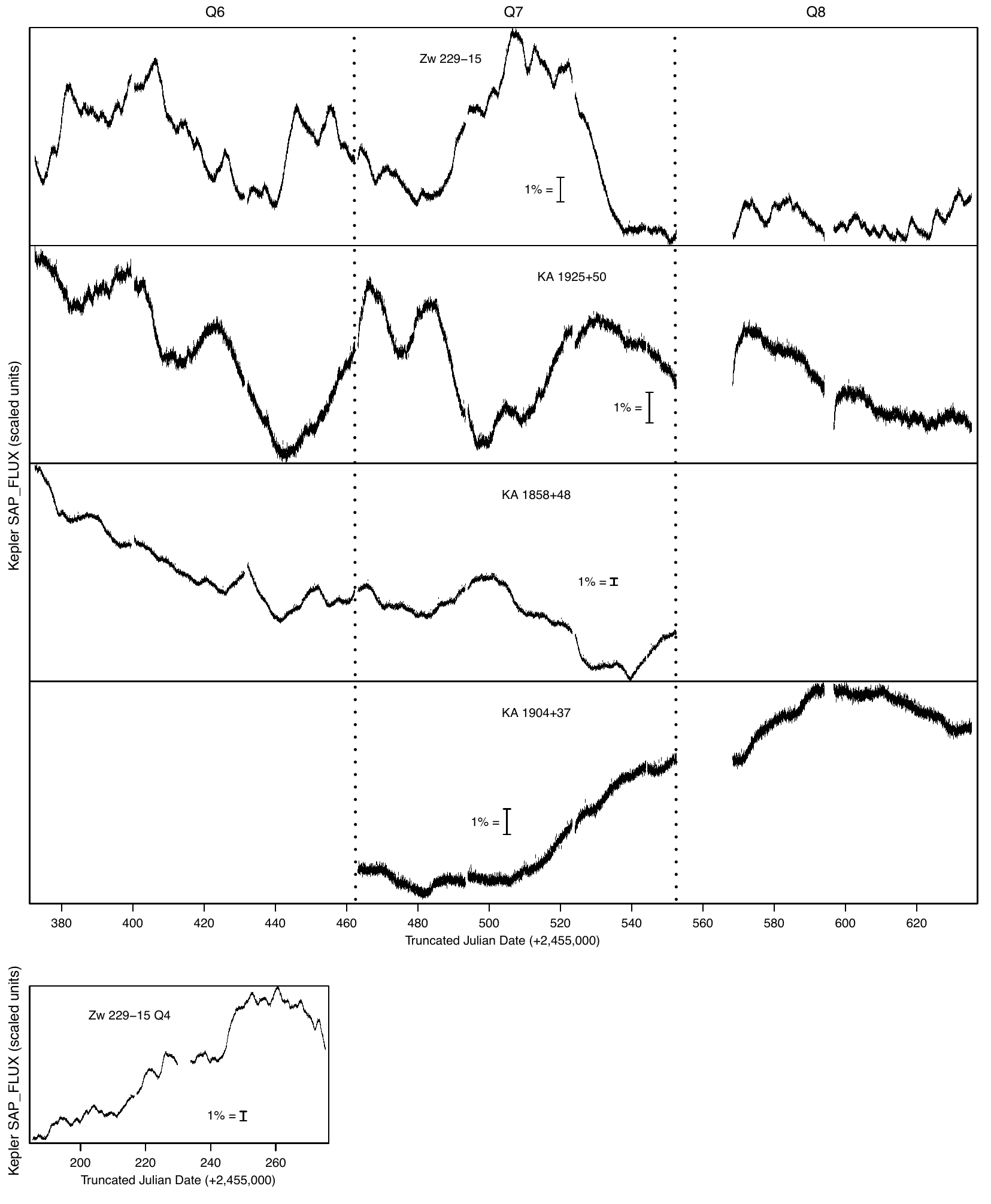}}
    \else
      \resizebox{\textwidth}{!}{\includegraphics{fig2.eps}}
    \fi
  \end{center}
\caption{Q6--Q8 light curves for four variable Kepler AGN.  A 1\% bar
  is shown for scale.  Q8 data were not obtained for KA 1858+48
  because it fell on defective Module 3.  Kepler observations of KA
  1904+37 did not begin until Q7.  Arbitrary offsets have been applied
  to match light curves across quarterly transitions (the dotted lines
  at TJD $\sim$ 462 and 552).  Note the 16 day gap due to a safe mode
  event at the beginning of Q8; this makes the offset for that quarter
  highly uncertain.  Note also that light curves occasionally show
  $\sim$1\% discontinuities immediately following monthly data
  downloads or safe mode events (e.g., TJD $\sim$ 568 and 586 in KA
  1925+50, and TJD $\sim$ 432 in Zw~229$-$15 and KA 1858+48) due to
  thermally induced focus changes.\\ The small bottom panel is the
  same as Figure 2a but for the Zw~229$-$15 Q4 data.
\label{figure2}}
\end{figure*}

The Kepler standard data processing pipeline (Jenkins et al. 2010),
operates on original spacecraft data to produce calibrated pixel data
(Quintana et al. 2011).  The next step, \texttt{PA}, uses simple
aperture photometry to extract \sapflux\ count rates from these
2-dimensional images (Twicken et al. 2011). The spacecraft downloads
not full CCD frames but only ``postage stamp'' images for the targets.
Only a fraction of the downloaded pixels are used in the extraction.
The next step in the standard pipeline, \sappdc, conditions the light
curves for transit searches, outputting \pdcflux\ light curves.
However, no conditioning occurred for sources presented in this paper
(the \sapflux\ and \pdcflux\ data are identical to within a constant
offset), so this and all further steps are not relevant to the current
work.  We use \sapflux\ count rates for our AGN light curve analyses.
These light curves are presented in Figures~\ref{figure1} and
\ref{figure2}.

Kepler, with its $ \lesssim 0.1$\% repeatability, $>90$\% duty cycle
and durations of years, explores a level of data quality superior to
anything previously obtained.  Thus one must be concerned about other
sources of error, especially systematic errors, in this relatively
young mission. An independent check of the Kepler data is available
for Zw229$-$15 since in 2010, it was observed by both Kepler and the
ground based Lick AGN Monitoring Program (LAMP).  These light curves,
shown in Figure 1, indicate a very good agreement between Kepler and
independent ground based LAMP data, well within the LAMP $\sim$1\%
errors and so, at least in this case, the systematic and other errors
in the Zw~229$-$15 data are generally no larger than the $\sim$1\% LAMP
errors.

However, the quoted Kepler errors are much smaller, and there is
currently no way to be sure that systematic errors are not affecting
the data at the level between $\sim$0.1\% and $\sim$1\%.  Indeed,
Figure 2 shows that small, short term (1--2 day), discontinuities are
sometimes observed following monthly data downloads or safe mode
events.  This is believed to arise from thermally induced focus
changes as the solar illumination changes during spacecraft
slews\footnote{\texttt{http://archive.stsci.edu/kepler/release\_notes/
    release\_notes5/Data\_Release\_05\_2010060414.pdf}}.  Both our
group and the Kepler team are working to correct for this in future
analyses.  While our understanding will undoubtedly improve as the
mission progresses, all that can be done at this time is to remind the
reader that systematic errors of this sort could still be present in
these data.

\section{Power Spectral Density Functions}

\subsection{PSD Measurement}

The optical flux variations in AGN are aperiodic. A standard tool for
characterizing such broadband (in temporal frequency) variability is
the periodigram, which measures the fluctuation power spectral density
(PSD) function. AGN PSDs have been best studied in the X-rays, where
the PSDs show a broad shape that has been simply characterized as a
double power law that breaks from a steep red noise high frequency
slope of $\alpha_H \sim -2$ ($S \propto f^\alpha$, where $\alpha$ is
the slope, $S$ is the spectral density and $f$ is the temporal
frequency) to a flatter low frequency slope of $\alpha_L \sim -1$, at
a break frequency $f_b$ that typically corresponds to timescales of
order a week, but scales with the mass of the black hole (e.g.,
Edelson \& Nandra 1999, Uttley et al. 2002, Markowitz et al. 2003).

We used the Kepler SAP data to measure PSDs for all of these Kepler
AGN.  Currently, large photometric offsets introduced by quarterly
spacecraft rolls prevent data from being combined across quarters, so
these PSDs only cover individual quarters.  This problem should
eventually be solved, so we will produce PSDs covering longer
timescales in a future paper.

For each light curve, a first order function was subtracted off so
that the first and last points of the light curve were equal.  This
``end-matching'' removes spurious low frequency power introduced by the
cyclic nature of the PSD which tends to flatten the PSDs. (See Fougere
1985 for details.)  This correction steepens the slopes by a mean
value of 0.7, 0.3, 0.8 and 0.7 for Zw~229$-$15, KA 1925+50, KA 1858+48
and KA 1904+37, respectively.  Fractional normalization was used, so
the resulting power density has units of rms$^2$~Hz$^{-1}$.

\begin{figure}
  \begin{center}
    \ifpdf
      \resizebox{\columnwidth}{!}{\includegraphics{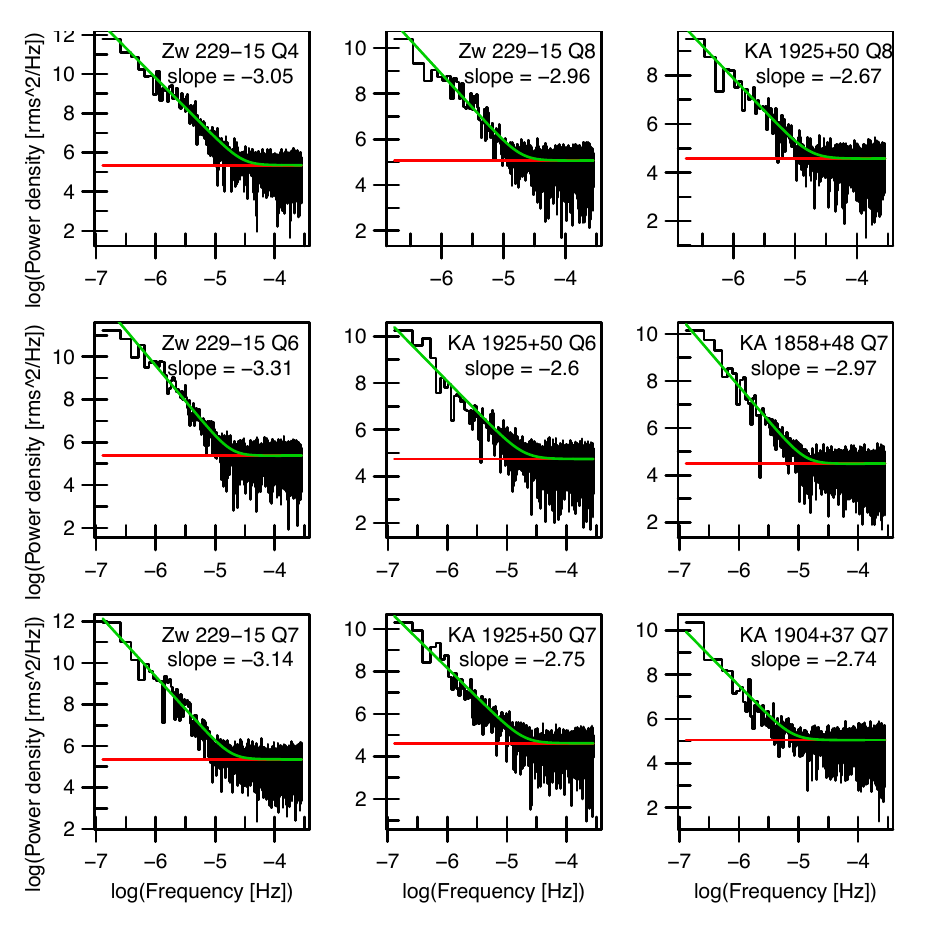}}
    \else
      \resizebox{\columnwidth}{!}{\includegraphics{fig3.eps}}
    \fi
  \end{center}
\caption{Optical PSDs and power law plus white noise fits for the 4
  AGN in selected quarters over temporal frequencies $\sim 10^{-6.5}$
  to $10^{-3.5}$~Hz.  The fits are shown in green, and the noise level
  in red.  Source name, quarter, and fitted power law slope ($\alpha$)
  are given in the upper right of each plot.
\label{figure3}}
\end{figure}

The resulting PSDs (see Figure~\ref{figure3}), fitted with a single
power law ($S \propto f^\alpha$) plus noise model on temporal
frequencies of $\sim4\times10^{-7}$ to $\sim 4\times 10^{-5}$~Hz
(corresponding to timescales of $\sim$6~hours to $\sim$1~month), are
very steep with slopes from $\alpha = -2.6$ to $-3.3$.

\subsection{Error analysis}

These PSDs also allow a check of the true noise level in the light
curves.  The fractional error, $err_{dir} = \langle err \rangle /
\langle flux \rangle$, is reported in Column 4 of Table~\ref{table2}.
An independent method of determining the error from the PSD uses the
formula of Vaughan et al. (2003): $err_{ind} = \sqrt{\langle err^2
  \rangle /\langle flux \rangle ^2}$, and is given in Column 5.  This
reduces to the same quantity $\left(\langle err\rangle /\langle
flux\rangle \right)$ in the limit of small fluctuations in the fluxes
and errors, as is the case with these data.  The errors derived from
the PSD analysis are typically $\sim25$\% larger than the quoted light
curve errors.  This indicates the quoted errors are slightly
underestimated, and that no other source of systematics dominates the
quoted errors.

\begin{deluxetable}{lccccc}
\tablecaption{Kepler AGN Observations \label{table2}}
\tablewidth{\columnwidth}
\tablehead{
\colhead{Source Name} &\colhead{Quarter} &\colhead{$10^3$ cts s$^{-1}$} 
&\colhead{err\_dir} &\colhead{err\_ind} &\colhead{$\alpha$}}
\startdata
Zw 229$-$15& Q4 & 12.1 & 0.047\% & 0.065\% & $-$3.05\\
Zw 229$-$15 & Q6 & 12.0 & 0.051\% & 0.068\% & $-$3.31\\
Zw 229$-$15 & Q7 & 12.9 & 0.046\% & 0.062\% & $-$3.14\\
Zw 229$-$15 & Q8 & 10.4 & 0.052\% & 0.055\% & $-$2.96\\
\hline
KA 1925+50& Q6 & 4.2 & 0.071\% & 0.084\% & $-$2.60\\
KA 1925+50 & Q7 & 3.8 & 0.065\% & 0.081\% & $-$2.75\\
KA 1925+50 & Q8 & 4.1 & 0.075\% & 0.078\% & $-$2.67\\
\hline
KA 1858+48 & Q6 & 2.1 & 0.117\% & 0.159\% & $-$2.87\\
KA 1858+48 & Q7 & 1.3 & 0.128\% & 0.207\% & $-$2.97\\
\hline
KA 1904+37 & Q7 & 5.8 & 0.071\% & 0.097\% & $-$2.74\\
KA 1904+37 & Q8 & 5.5 & 0.080\% & 0.087\% & $-$2.95
\enddata
\tablecomments{Columns 1 and 2 give the source name and quarter,
  column 3 the mean \sapflux\ count rate in units of
  $10^3$~cts~s$^{-1}$, and column 4 the ratio of the mean quoted
  errors divided by the mean flux.  Column 5 gives the error rate
  derived from the PSD fits as discussed in Section 3.2.  Column 6
  gives the fitted PSD slopes ($\alpha$) for each quarter.}
\end{deluxetable}

The PSD slopes for each quarter (listed in Table~\ref{table2}) show
small scatter for individual objects. It is difficult to directly
measure reliable errors on derived PSD slopes, but an estimate is
provided by the observed dispersion for individual objects.  For the
two sources with the most data, Zw~229$-$15 and KA~$1925+50$, the mean
slope and associated standard deviations are $\langle \alpha\rangle =
-3.11 \pm 0.15$ and $-2.67 \pm 0.08$.  These differ by $\sim$2.5
standard deviations, suggesting, at very marginal significance, that
the intrinsic difference between the derived slopes for these objects
is larger than the associated errors.  (The quoted uncertainties are
standard deviations of the distributions of the PSD slopes for
different quarters.) Note that without the red noise leak correction,
the standard deviations for these two sources would have been 0.58 and
0.22, respectively, so our correction successfully reproduces similar
PSD slopes between the various quarters for each source. Since PSD
analyses are notoriously susceptible to analytical systematics (see
e.g., Vaughan et al. 2003) and there is the possibility that currently
unknown systematic errors could affect these new Kepler data (see
Sect. 2.2), the agreement in slope from quarter to quarter provides a
degree of confidence that the observed steep slopes are accurate.

\section{Discussion}

\subsection{Comparison to Previous Results}

\subsubsection{Optical Data}

Kepler light curves are of much higher quality and sampling rate than
previous data. For example: in the data used by Kelley et al. (2009)
the highest photometric quality is from the MACHO survey of Geha et
al. (2003) which has $\sim$5\% photometric errors and 600 good
photometric measurements over 7.5~years, and thus samples at $\sim$1
point every 4.5~days compared to the 0.1\% Kepler errors and 1 data
point roughly every 30~minutes. Previous attempts to derive the PSD
over a wide range of timescales have had to combine the data from many
objects and several surveys (Hawkins 2002) or have relied on
relatively sparsely sampled data, from several different telescopes
(Breedt et al. 2010).

Previous results (e.g. Kelly et al 2009) tend to find best fitting
PSDs with slopes of $\sim-1.8$ for the collective sample, rather
flatter than what we have found. Since the Kepler PSDs cannot continue
to very low frequencies with such steep slopes without implying very
large variability amplitudes, there must be a break at timescales
$>1$~month, which may make the Kepler PSDs consistent with previous
work. It is not surprising that the results of our observations are
rather different than what has been published previously---the other
observations could not see the effects we are detecting. While there
is a formal overlap in sampled timescales between our Kepler and other
data, the much larger error bars for the previous PSDs (e.g. Breedt et
al. 2010,) at characteristic frequencies above a few $\times
10^{-5}$~Hz makes comparison difficult. However, for at least one
object, NGC~4051 (Breedt et al 2010), the observed PSD in the
$10^{-6}$-- $10^{-8}$~Hz range is well determined and is flatter than
our Kepler results for all of our objects. One possible explanation
for the differences may lie in the different luminosities or Eddington
ratios of the objects, since NGC~4051 is significantly less luminous
and probably less massive than the objects in our sample.

\subsubsection{X-ray Data}

Although the particular Seyfert~1s in our sample do not have measured
X-ray PSDs, many other Seyfert~1s have had X-ray PSDs measured over
these timescales. These are always much flatter, typically having high
frequency slopes of $-1$ to $-2$ (Edelson \& Nandra 1999, Uttley et
al. 2002, Markowitz et al. 2003).  Thus our measurement of steep
optical PSDs on short timescales is somewhat surprising because it is
so different from that measured in the X-rays, and because Seyfert~1
optical and X-ray light curves appear to track well, at least on
longer timescales (Uttley et al. 2003).

\subsection{Physical Implications for Accretion Disks}

The characteristic timescales of the fluctuations should correspond to
different physical mechanisms which may be related to the size of the
system, the dynamical timescales, epicyclic frequencies, g-modes or
other characteristic timescales which could influence the source of
variance.  Since the source of the accreting material in AGN is not
known, it is unclear if the sources of the perturbations are changes
in the accretion flow, the turbulence due to physics in the disk
itself (from the magnetorotational instability mechanism (MRI),
e.g. Miller \& Reynolds 2009, Noble \& Krolik 2009), or perhaps other
physics. As shown by McHardy et al. (2006), the characteristic
timescale seen in the X-ray PSDs is related to the AGN mass and the
accretion rate.  However, it is not known if this is also true for the
optical PSDs (MacLeod et al. 2010).

Recent results from ground based optical observations (e.g. Kelly et
al. 2009, MacLeod et al. 2010) find that their results are consistent
with a ``damped random walk model''.  However, their light curves are
irregularly and more sparsely sampled compared to Kepler data (see
Figure 2 in Kozlowski et al. 2010).  Our data do not find the
predicted $f^{-2}$ power spectrum at high frequencies predicted by
this model. However, since there is very little overlap in frequencies
and our sample size is much smaller, direct comparison is
difficult. Our data are just capable of reaching the light travel time
size of the disks on our sampled AGN.  The effective size of the
region emitting radiation at a given frequency is (Baganoff \& Malkan
1995):
\[ r_{1/2} = 7.5 \times 10^{23} \epsilon^{-1/3} \nu^{-4/3}
\,(\textrm{M/\msun})^{-1/3}\,(\textrm{L/L}_{\textrm{Edd}})^{1/3}\,r_G,\]
where $r_G$ is the Schwarzschild radius, $\epsilon$ is the accretion
efficiency and $\nu$ is the effective observing frequency of the data.
Utilizing an effective wavelength of 5000\AA, mass of
$1\times10^7$~\msun\ (Barth et al 2011) and Eddington ratio of 0.05 we
find an effective light travel time ($r_{1/2}/c$) of $\sim$1 day which
is close to our white noise limit of 0.25~days. The 4 sources in this
paper span only 1 order of magnitude in X-ray luminosity ($\log L_X =
42.6$~to~43.6) and thus, probably, a small range in mass. Our future
observations we will have a larger number as well as more luminous
objects and thus we should constrain the limits where light travel
time effects can be well measured.

While modeling of accretion disks from first principles via
magnetohydrodynamic (MHD) calculations is in its early days there are
several estimates of the slope of the PSD from accretion disks.  In
these models the underlying physical drivers for variability in the
light curve are variations in the accretion rate caused by the chaotic
character of MHD turbulence. Noble and Krolik~(2009) simulate emission
from the coronae appropriate to the X-ray emission, and thus it is not
clear if their simulation is comparable with our results.  Chan et
al. (2009) focus on Sgr~A* which seems to be accreting in a different
mode than the Seyfert~1s in our sample.  Reynolds \& Miller (2009)
show PSDs of the mass accretion rate whose high frequency slopes
($\sim -2.9$) are very close to those seen in our observations.
However, their simulation was only run for a relatively short time
($\sim1.2\times10^4$~$GM/c^3$) which corresponds to 14 days for
objects of the mass of Zw~229$-$15.

All simulations so far suffer from the fundamental problem that to
compare them with observations one has to convert the simulated disk
characteristics into a radiation flux spectrum. Thus it is not clear
that the proxies for emission developed so far are appropriate.  This
problem is fully recognized by the simulators and thus, in general,
they have been loath to directly compare to the data.

\section{Conclusions}

Power spectral analysis of four AGN observed by Kepler during Q6-Q8
show very steep ($\alpha \sim -2.6$ to $-3.3$) slopes, considerably
steeper than that seen in the X-rays. The PSDs for each source are
consistent from quarter to quarter and, at $>2\sigma$ confidence, are
different from each other.  Analysis of these high quality light
curves indicates that the influence of systematic errors is rather
small; additionally, direct comparison of Kepler and LAMP monitoring
of Zw~229$-$15 shows excellent agreement. Comparison with analytic
models of AGN variability shows steeper than predicted slopes;
however, comparison with MHD simulations seems to show better
agreement.  Further analysis of other characteristics of the light
curve, longer time series, the analysis of more objects and the
comparison to semi-analytic models of time variability will be the
subject of future papers. We hope that these new high quality Kepler
data will stimulate the calculation of the time series from accretion
disks.

\acknowledgements

We thank the Kepler team for their efforts to make the data accessible
and tractable and the Kepler GO program for funding, Matt Malkan for
extensive contributions to the identification of new Kepler AGN, Simon
Vaughan for valuable help with PSD measurements and Aaron Barth and
the LAMP team for early access to their data.

\end{document}